\documentstyle[amsfonts,aps]{revtex}

\begin{document}
\author{Mario Castagnino}
\address{CONICET-UNR-UBA, Institutos de F\'{i}sica de Rosario y de Astronom\'{i}a y\\
F\'{i}sica del Espacio.\\
Casilla de Correos 67, Sucursal 28, 1428, Buenos Aires, Argentina}
\author{Olimpia Lombardi}
\address{CONICET-IEC, Universidad Nacional de Quilmes\\
Rivadavia 2328, 6${{}^o}$ Derecha, 1034, Buenos Aires, Argentina.}
\title{Decoherence time in self-induced decoherence}
\maketitle

\begin{abstract}
A general method for obtaining the decoherence time in self-induced
decoherence is presented. In particular, it is shown that such a time can be
computed from the poles of the resolvent or of the initial conditions in the
complex extension of the Hamiltonian's spectrum. Several decoherence times
are estimated: $10^{-13}-$ $10^{-15}s$ for microscopic systems, and $%
10^{-37}-10^{-39}s$ for macroscopic bodies. For the particular case of a
thermal bath, our results agree with those obtained by the einselection
(environment-induced decoherence) approach.

PACS number(s) 03.65.Bz

e-mail: mariocastagnino@citynet.net.ar
\end{abstract}

\section{Introduction}

The phenomenon of decoherence is usually considered as a relevant element
for understanding how the classical world emerges from an underlying quantum
realm. In a first period, decoherence was explained as the result of the
destructive interference of the off-diagonal elements of a density matrix
(see \cite{Van Kampen}, \cite{Daneri et al.}); however, this line of
research was abandoned due to technical difficulties derived from the
formalism used to describe the process. As a consequence, decoherence begun
to be conceived as produced by the interaction between a system and its
environment. This approach gave rise to the {\it environment-induced
decoherence} program, based on the works of Zeh (\cite{Zeh 1970}, \cite{Zeh
1971}, \cite{Zeh 1973}) and later developed by Zurek and coworkers (\cite
{Zurek 1981}, \cite{Zurek 1982}, \cite{Zurek 1991}, \cite{Zurek 1993}, \cite
{Zurek 1998}, \cite{Paz}, \cite{Zurek 2003}). Although many relevant results
have been obtained by means of environment-induced decoherence, this
approach still involves certain unsolved problems (see \cite{SHPMP}, \cite
{Max-1}):

\begin{enumerate}
\item  Einselection is based on the decomposition of the system into a
relevant part, the {\it proper system,} and an irrelevant part, the {\it %
environment.} This decomposition is not always possible, as in the case of
the universe. In fact, Zurek himself considers the criticism: ''...{\it \
the Universe as a whole is still a single entity with no 'outside'
environment, and therefore any resolution involving its division is
unacceptable}'' (\cite{Zurek 1994}, p.181). The same problem appears in any
closed system (and, therefore, with no interaction with an environment) that
becomes classical. In fact, if ''{\it ...the existence of emergent
classically will be always accompanied by other manifestations of openness
such a dissipation of energy into the environment}'' (\cite{Paz}, p.6), the
problem is to explain why many systems behave classically maintaining their
energy constant.

\item  The einselection approach does not provide a clear definition of the
'cut' between the proper system and its environment. In fact, as Zurek
himself admits, ''{\it In particular, one issue which has been often taken
for granted is looming big, as a foundation of the whole decoherence
problem. It is the question of what the 'system' is which plays such a
crucial role in all the discussions of the emergent classicality. This
question was raised earlier but the progress to date has been slow at best}%
'' (\cite{Zurek existential} p.122).

\item  In the einselection approach, the definition of the basis where the
system becomes classical, i.e. the {\it pointer basis}, relies on the
'predictability sieve' which would produce the set of the most stable
states. But this definition seems very difficult to implement, at least in a
generic case. In fact, the basis vectors are only good candidates{\it \ }for
reasonable stable states (see \cite{Zurek Los Alamos}).
\end{enumerate}

As the result of these and other difficulties, a number of alternative
accounts of decoherence have been proposed (see \cite{Casati-1}, \cite
{Casati-2}, \cite{Penrose}, \cite{Diosi-1}, \cite{Diosi-2}, \cite{Milburn}, 
\cite{Adler}).

In a series of papers (\cite{SHPMP}, \cite{Cast-Gadella 1997}, \cite
{Cast-Laura 1997}, \cite{Laura-Cast 1998-E}, \cite{Laura-Cast 1998-A}, \cite
{Cast 1999}, \cite{Cast-Laura 2000-PRA}, \cite{Cast-Laura 2000-IJTP}, \cite
{Cast-Gad-Laura-Betan 2001-PLA}, \cite{Cast-Gad-Laura-Betan 2001-JPA}, \cite
{Cast-Lombardi 2003}, \cite{Cast-Gadella 2003}, \cite{Cast-Ordoñez 2004}, 
\cite{Cast 2004}, \cite{Cast-Lombardi 2005}) we have returned to the initial
idea of the destructive interference of the off-diagonal terms of the
density matrix, but now on the basis of the van Hove formalism (\cite{van
Hove 1955}, \cite{van Hove 1956}, \cite{van Hove 1957}, \cite{van Hove 1959}%
, \cite{van Hove 1979}). We have called this new approach '{\it self-induced
decoherence}' because, from this viewpoint, decoherence is not produced by
the interaction between a system and its environment, but results from the
own dynamics of the whole quantum system governed by a Hamiltonian with
continuous spectrum. The aim of this paper is to present a general method
for obtaining the decoherence time in self-induced decoherence. In
particular, we will show that such a time can be computed from the poles of
the resolvent or of the initial conditions in the complex extension of the
Hamiltonian's spectrum. The general formalism has been developed in papers 
\cite{Rolo} and \cite{Liotta}, but in the context of a discussion about the
nature and properties of Gamow vectors. Here we will adopt that formalism
for the computation of the decoherence time.

The paper is organized as follows. In Section II we briefly review the
formal basis of self-induced decoherence. In Section III we show that the
decoherence time can be computed as the characteristic decaying time of the
expectation value equation, and that this time can be obtained in terms of
the poles of the involved functions. In Section IV we find the origin of
these poles in the resolvent of the Hamiltonian or in the initial
conditions. In Sections V and VI we estimate the decoherence time in
microscopic systems and in macroscopic bodies, respectively. Section VII is
devoted to study the case of a thermal bath, where the decoherence time is
estimated and compared with the corresponding results obtained by the
einselection approach. After the conclusions, we include two appendices: in
Appendix A we introduce a necessary mathematical remark, and in Appendix B
we sketch a model with two characteristic times: decoherence and relaxation.

\section{Self-induced decoherence}

In this section we will present the formalism of self-induced decoherence by
means of a very simple case. We refer the reader to our previous papers; in
particular, for more complex cases, see \cite{Cast-Laura 2000-PRA}, and for
a conceptual discussion about the physical meaning of self-induced
decoherence, see \cite{SHPMP}.

Let us consider a quantum system endowed with a Hamiltonian with continuous
spectrum 
\begin{equation}
H=\int_{0}^{\infty }\omega |\omega \rangle \langle \omega |d\omega
\label{2-1}
\end{equation}
A generic observable reads 
\begin{equation}
O=\int_{0}^{\infty }\int_{0}^{\infty }\widetilde{O}(\omega ,\omega ^{\prime
})|\omega \rangle \langle \omega ^{\prime }|d\omega d\omega ^{\prime }
\label{2-2}
\end{equation}
where $O(\omega ,\omega ^{\prime })$ is any kernel or distribution. Since
the algebra of these observables is too large for our purposes, we only
consider the van Hove operators such that 
\begin{equation}
\widetilde{O}(\omega ,\omega ^{\prime })=O(\omega )\delta (\omega -\omega
^{\prime })+O(\omega ,\omega ^{\prime })  \label{2-3}
\end{equation}
where $O(\omega ,\omega ^{\prime })$ is a regular function. Then, 
\begin{equation}
O=\int_{0}^{\infty }O(\omega )|\omega \rangle \langle \omega |d\omega
+\int_{0}^{\infty }\int_{0}^{\infty }O(\omega ,\omega ^{\prime })|\omega
\rangle \langle \omega ^{\prime }|d\omega d\omega ^{\prime }  \label{2-4}
\end{equation}
where the first term of the r.h.s. is the {\it singular term }$O_{S}${\it ,}
and the second term is the {\it regular term }$O_{R}$. These observables
belong to an algebra $\widehat{{\cal A}}$, that we will call '{\it van Hove
algebra}': they are the self-adjoint operators of $\widehat{{\cal A}}$ and
belong to a space of operators $\widehat{{\cal O}}$. The basis of $\widehat{%
{\cal A}}$ is $\{|\omega ),|\omega ,\omega ^{\prime })\}$, where $|\omega
)=|\omega \rangle \langle \omega |$ and $|\omega ,\omega ^{\prime })=|\omega
\rangle \langle \omega ^{\prime }|$.

Let us now consider the space $\widehat{{\cal O}^{\prime }}$, that is, the
dual space $\widehat{{\cal O}}$ with basis $\{(\omega |,(\omega ,\omega
^{\prime }|\}$. A generic state belonging to $\widehat{{\cal O}^{\prime }}$
reads 
\begin{equation}
\rho =\int_{0}^{\infty }\rho (\omega )(\omega |d\omega +\int_{0}^{\infty
}\int_{0}^{\infty }\rho (\omega ,\omega ^{\prime })(\omega ,\omega ^{\prime
}|d\omega d\omega ^{\prime }  \label{2-5}
\end{equation}
and must satisfy the usual constraints: $\rho (\omega )$ must be real and
positive and $\int_{0}^{\infty }\rho (\omega )d\omega =1$. We also introduce
the requirement that $\rho (\omega ,\omega ^{\prime })$ be a regular
function. Again, the first term of the r.h.s. of eq.(\ref{2-5}) is the {\it %
singular term} $\rho _{S}$, and the second term is the {\it regular term} $%
\rho _{R}$.

The expectation value of the observable $O$ in the state $\rho $ results
from the action of the functional $\rho $ on the operator $O$, $(\rho |O)$, 
\begin{equation}
\langle O\rangle _{\rho }=\int_{0}^{\infty }\overline{\rho (\omega )}%
O(\omega )d\omega +\int_{0}^{\infty }\int_{0}^{\infty }\overline{\rho
(\omega ,\omega ^{\prime })}O(\omega ,\omega ^{\prime })d\omega d\omega
^{\prime }  \label{2-6}
\end{equation}
where functions $\overline{\rho (\omega )}$ and $O(\omega )$ are such that
the first integral is well defined. The time evolution of this last equation
reads 
\begin{equation}
\langle O\rangle _{\rho (t)}=\int_{0}^{\infty }\overline{\rho (\omega )}%
O(\omega )d\omega +\int_{0}^{\infty }\int_{0}^{\infty }\overline{\rho
(\omega ,\omega ^{\prime })}O(\omega ,\omega ^{\prime })\,e^{i\frac{\omega
-\omega ^{\prime }}{\hbar }t}d\omega d\omega ^{\prime }  \label{2-7}
\end{equation}
Since the first term of the r.h.s. is time-constant and the second is a
function of time, we will call them '{\it constant term}' and '{\it %
fluctuating term}', respectively.

The Riemann-Lebesgue theorem, which mathematically expresses the phenomenon
of destructive interference, states that, if $f(\nu )\in {\Bbb L}_{1}$, 
\begin{equation}
\lim_{t\rightarrow \infty }\int d\nu f(\nu )e^{i\nu t}=0  \label{2-8}
\end{equation}
Therefore, if the function $\overline{\rho (\omega ,\omega ^{\prime })}%
O(\omega ,\omega ^{\prime })$ of eq.(\ref{2-7}) is ${\Bbb L}_{1}$ in
variable $\nu =\omega -\omega ^{\prime }$, we can apply the Riemann-Lebesgue
theorem, 
\begin{equation}
\lim_{t\rightarrow \infty }(\rho _{R}(t)|O_{R})=\lim_{t\rightarrow \infty
}\int_{0}^{\infty }\int_{0}^{\infty }\overline{\rho (\omega ,\omega ^{\prime
})}O(\omega ,\omega ^{\prime })\,e^{i\frac{\omega -\omega ^{\prime }}{\hbar }%
t}d\omega d\omega ^{\prime }=0  \label{2-9}
\end{equation}
As a consequence, 
\begin{equation}
\lim_{t\rightarrow \infty }\langle O\rangle _{\rho (t)}=\lim_{t\rightarrow
\infty }(\rho (t)|O)=\int_{0}^{\infty }\overline{\rho (\omega )}O(\omega
)d\omega  \label{2-10}
\end{equation}
This last equation can be expressed as a weak-limit 
\begin{equation}
W-\lim_{t\rightarrow \infty }(\rho (t)|=(\rho _{*}|=\int_{0}^{\infty }%
\overline{\rho (\omega )}(\omega |d\omega  \label{2-11}
\end{equation}
where $(\rho _{*}|$ has only singular-diagonal terms.\footnote{%
Here we are working in the Schr\"{o}dinger picture. However, analog results
can be obtained in the Heisenberg picture: 
\[
\lim_{t\rightarrow \infty }\langle O(t)\rangle _{\rho }=\lim_{t\rightarrow
\infty }(\rho |O(t))=\int_{0}^{\infty }\overline{\rho (\omega )}O(\omega
)d\omega =(\rho |O_{*}) 
\]
}

It is not difficult to see that this approach to decoherence avoids the
problems of the einselection program. In fact, self-induced decoherence can
be applied to closed systems as the universe \cite{Cast-Lombardi 2003}, the
problem of the 'cut' between the proper system and its environment is absent 
\cite{SHPMP}, and the pointer basis is perfectly defined (e.g. the energy
eigenbasis in the above example) \cite{Laura-Cast 1998-A}. Nevertheless, two
points need to be emphasized:

\begin{enumerate}
\item  In this approach, the coarse graining necessary to turn the usual
unitary time evolutions of quantum mechanics into non-unitary time
evolutions is not explicit in the formalism. However, the choice of a
particular algebra $\widehat{{\cal A}}$ , the van Hove algebra, among all
possible algebras and the systematic use of expectation values $\langle
O\rangle _{\rho (t)}=(\rho (t)|O)$ play the role of a coarse graining. In
fact, we can define the projector $\Pi =|O)(\rho _{0}|$, with $|O)\in 
\widehat{{\cal A}}$ and $(\rho _{0}|O)=1$, that projects $(\rho (t)|$ as $%
(\rho (t)|\Pi =\langle O\rangle _{\rho (t)}(\widehat{\rho }_{0}|$ and allows
us to translate our results in the language of projectors: in this case,
from eq.(\ref{2-10}) we will obtain $\lim_{t\rightarrow \infty }(\rho
(t)|\Pi =(\rho _{*}|\Pi $, where $(\rho _{*}|$ is diagonal. This projection
is what breaks the unitarity of the primitive evolution. A detailed
discussion on this point can be found in \cite{SHPMP} and \cite
{Cast-Lombardi 2005}.

\item  As a consequence of Riemann-Lebesgue theorem, full decoherence occurs
at $t\rightarrow \infty $. However, as in any exponential decaying process,
there is a characteristic decaying time that can be considered as the time
at which, in practice, the decaying is approximately completed. In the next
sections we will compute the decoherence time of a self-induced decoherence
process as the characteristic decaying time of the fluctuating term in the
expression of $\langle O\rangle _{\rho (t)}$ of eq.(\ref{2-7}).
\end{enumerate}

\section{Poles in the expectation value equation}

In order to study and compute the decoherence time, we will use the standard
theory of analytical continuation in scattering quantum theory (see e.g. 
\cite{Arno}, \cite{Ballentine}) and its extension to the Liouville-von
Neumann space (see e.g. \cite{Cast-Laura 1997} and \cite{Rolo}). By means of
this theory, we can compute the decoherence time in terms of the poles
corresponding to the functions involved in the fluctuating term of $\langle
O\rangle _{\rho (t)}$ (see eq.(\ref{2-7})): 
\begin{equation}
(\rho _{R}(t)|O_{R})=\int_{0}^{\infty }\int_{0}^{\infty }\overline{\rho
(\omega ,\omega ^{\prime })}O(\omega ,\omega ^{\prime })\,e^{i\frac{\omega
-\omega ^{\prime }}{\hbar }t}d\omega d\omega ^{\prime }  \label{3-1}
\end{equation}

In this equation, we can introduce the following change of variables (see
Appendix A): 
\begin{equation}
\lambda =\frac{1}{2}(\omega +\omega ^{\prime }),\qquad \nu =\omega -\omega
^{\prime },\qquad d\omega d\omega ^{\prime }=Jd\lambda d\nu =d\lambda d\nu
\label{3-2}
\end{equation}
Then, 
\begin{equation}
(\rho _{R}(t)|O_{R})=\int_{0}^{\infty }d\lambda \int_{-2\lambda }^{2\lambda
}d\nu \,\overline{\rho ^{\prime }(\nu ,\lambda )}O^{\prime }(\nu ,\lambda
)\,e^{i\frac{\nu }{\hbar }t}\qquad  \label{3-3}
\end{equation}
where $\overline{\rho ^{\prime }(\nu ,\lambda )}=\overline{\rho (\omega
,\omega ^{\prime })}$, $O^{\prime }(\nu ,\lambda )=O(\omega ,\omega ^{\prime
})$ and the new limits of the integrals are due to the fact that $\omega
,\omega ^{\prime }\geq 0$. Now we promote the real variable $\nu $ to a
complex variable $Z$; if the function $\overline{\rho ^{\prime }(Z,\lambda )}%
O^{\prime }(Z,\lambda )$ has {\it no poles} in the upper $Z-$half-plane, we
obtain 
\begin{equation}
(\rho _{R}(t)|O_{R})=\int_{0}^{\infty }d\lambda \int_{C(-2\lambda ,2\lambda
)}dz\,\overline{\rho ^{\prime }(Z,\lambda )}O^{\prime }(Z,\lambda )\,e^{i%
\frac{Z}{\hbar }t}\qquad  \label{3-4}
\end{equation}
where $C(-2\lambda ,2\lambda )$ is any curve that goes from $-2\lambda $ to $%
2\lambda $ by the upper complex half-plane. If the function $\overline{\rho
^{\prime }(Z,\lambda )}O^{\prime }(Z,\lambda )$ has, say, a pole at $Z_{0}=%
\widetilde{\omega }+i\gamma $ in the upper half-plane, we can, as usual,
decompose $C(-2\lambda ,2\lambda )=\Gamma (-2\lambda ,2\lambda )\cup
C_{Z_{0}}$, where $C_{Z_{0}}$ is a residue-circle around the pole $Z_{0\text{
}}$and $\Gamma (-2\lambda ,2\lambda )$ is the remaining 'background' curve.
If, as usual, we neglect the background, only the factor $e^{i\frac{Z}{\hbar 
}t}$ becomes relevant at the pole $Z_{0}$; this factor reads 
\begin{equation}
e^{i\frac{Z_{o}}{\hbar }t}=e^{i\frac{\widetilde{\omega }+i\gamma }{\hbar }%
t}=e^{i\frac{\widetilde{\omega }}{\hbar }t}e^{-\frac{\gamma }{\hbar }t}
\label{3-5}
\end{equation}
where $e^{-\frac{\gamma }{\hbar }t}$ is a dumping factor appearing in the
regular fluctuating term of $\langle O\rangle _{\rho (t)}=(\rho (t)|O)$.
Therefore, the decoherence time can be computed as the characteristic
decaying time of the process as 
\begin{equation}
t_{D}=\frac{\hbar }{\gamma }  \label{3-6}
\end{equation}
Let us note that, up to this point, we have worked in the eigenbasis of the
complete Hamiltonian $H$, that we will call $\{|\omega \rangle ^{+}\}$.%
\footnote{%
Strictly speaking, we will find a double basis $\left\{ |\omega \rangle
^{\pm }\right\} $ corresponding to a decaying process and a growing process,
respectively. But since we are only interested in the former one, we will
use only $\left\{ |\omega \rangle ^{+}\right\} $ and work in the lower
half-plane, trying to find poles in the second sheet. In Section II, we have
simply called $|\omega \rangle $ the eigenvectors $|\omega \rangle ^{+}$
(see eq.(\ref{2-1})).}

Once the decoherence time has been computed, a single question remains: what
is the origin of the pole in the product $\overline{\rho ^{\prime
}(Z,\lambda )}O^{\prime }(Z,\lambda )$? We will address this problem in the
next section.

\section{The origin of the poles}

In the physical evolutions we are interested on, there are 'free' periods
with no decoherence (e.g. very long ones like the 'in' and 'out' periods
used to modelize a scattering process, or short periods but long enough to
fix the initial conditions for an 'interaction' period) and 'interaction'
periods where decoherence occurs. On this basis, we will first consider in
detail the 'free' period governed by a 'free' Hamiltonian $H_{0}$ (Case 1),
and then the 'interaction' period governed by a 'perturbed' Hamiltonian $%
H=H_{0}+V$ (Case 2).

{\bf Case 1}: Let us consider the 'free' case with a free Hamiltonian $H_{0}$%
, and call $\{|E\rangle \}$ the eigenbasis of $H_{0}$, where $0\leq E<\infty 
$. Decoherence is due to the vanishing of the fluctuating term of $\langle
O\rangle _{\rho (t)}$ (see eq.(\ref{3-1}) 
\begin{equation}
\lim_{t\rightarrow \infty }(\rho _{R}(t)|O_{R})=\lim_{t\rightarrow \infty
}\int_{0}^{\infty }\int_{0}^{\infty }\overline{\rho (E,E^{\prime })}%
O(E,E^{\prime })\,e^{i\frac{E-E^{\prime }}{\hbar }t}dEdE^{\prime }=0
\label{4-1}
\end{equation}
where $\rho (E,E^{\prime })=\langle E|\rho |E^{\prime }\rangle $ are the
coordinates of the state, and $O(E,E^{\prime })$ $=\langle E|O|E^{\prime
}\rangle $ are the coordinates of the considered observable, both in the
basis $\{|E\rangle \}$. Calling, as above, 
\[
\lambda _{0}=\frac{1}{2}(E+E^{\prime }),\qquad \nu _{0}=E-E^{\prime } 
\]
the limit reads 
\begin{equation}
\lim_{t\rightarrow \infty }(\rho _{R}(t)|O_{R})=\int_{0}^{\infty }d\lambda
_{0}\lim_{t\rightarrow \infty }\int_{-2\lambda _{0}}^{2\lambda _{0}}d\nu _{0}%
\overline{\,\rho (\lambda _{0}+\nu _{0}/2,\lambda _{0}-\nu _{0}/2)}O(\lambda
_{0}+\nu _{0}/2,\lambda _{0}-\nu _{0}/2)\,e^{i\frac{\nu _{0}}{\hbar }t}
\label{4-2}
\end{equation}
Proceeding in the same way as in the previous section, we arrive to an
expression similar to eq.(\ref{3-4}) 
\begin{equation}
\lim_{t\rightarrow \infty }(\rho _{R}(t)|O_{R})=\int_{0}^{\infty }d\lambda
_{0}\lim_{t\rightarrow \infty }\int_{C(-2\lambda _{0},2\lambda _{0})}dZ\,%
\overline{\,\rho (\lambda _{0}+Z/2,\lambda _{0}-Z/2)}O(\lambda
_{0}+Z/2,\lambda _{0}-Z/2)\,e^{i\frac{Z}{\hbar }t}  \label{4-3}
\end{equation}
where some poles could be found (see details in \cite{Cast-Laura 1997} and 
\cite{Rolo}). However, since this is a 'free' period, there should be no
decoherence, i.e. the decoherence time should be infinite. This means that
we have to adjust our theory to this physical fact by asking the following
conditions for the complex continuation of the coordinates $\rho
(E,E^{\prime })=\langle E|\rho |E^{\prime }\rangle $, $O(E,E^{\prime })$ $%
=\langle E|O|E^{\prime }\rangle $:

\begin{itemize}
\item  {\bf Condition 1}: In the 'free' eigenbasis $\{|E\rangle \}$, the
coordinates of the observable $O$, $O(\lambda _{0}+Z/2,\lambda _{0}-Z/2)$,
have no poles in the upper half plane. This is a natural requirement
because, if not, the observable $O$ would introduce by itself a finite
decoherence time {\it for any} {\it state} and {\it any evolution
Hamiltonian,} which is certainly not a physical situation.

\item  {\bf Condition 2}: In the 'free' eigenbasis $\{|E\rangle \}$, the
coordinates of the state $\rho $, $\rho (\lambda _{0}+Z/2,\lambda _{0}-Z/2)$%
, have no poles in the upper half plane.\footnote{%
Nevertheless, in Section VII we will see that, in some cases, there are
poles in the 'free' period, e.g. in the case of the evolution of a system
with a thermal bath.}
\end{itemize}

If these two conditions are satisfied, the function $\overline{\,\rho
(\lambda _{0}+Z/2,\lambda _{0}-Z/2)}O(\lambda _{0}+Z/2,\lambda _{0}-Z/2)\,$%
will have no poles in the upper half-plane. In this case, the decoherence
time is {\it infinite} and decoherence is only nominal, as one would have
expected in a free evolving situation.

{\bf Case 2}: Once we have 'calibrated' our theory in order to satisfy the
physical condition according to which a free evolving system does not
decohere, we will consider the process in the 'interaction' period. The
total Hamiltonian now reads

\begin{equation}
H=H_{0}+V=\int_{0}^{\infty }\omega |\omega \rangle \langle \omega |\,d\omega
+\int_{0}^{\infty }\int_{0}^{\infty }V(\omega ,\omega )|\omega \rangle
\langle \omega ^{\prime }|\,d\omega d\omega ^{\prime }  \label{4-4}
\end{equation}
where $\{|\omega \rangle \}$ is the eigenbasis of $H_{0}$ (here we have
replaced $E$ with $\omega $ to emphasize that now this is an 'interaction'
period). Let us consider the resolvent, namely, the {\it complex value
operator }(see \cite{Kato}) 
\begin{equation}
R(z)=(z-H)^{-1}  \label{4-5}
\end{equation}
i.e. the analytical continuation to the lower second sheet of 
\begin{equation}
R(\omega )=(\omega +i0-H)^{-1}  \label{4-6}
\end{equation}
The poles of $R(z)$ are known as the poles of the resolvent, and they
coincide with those of the S-matrix. In fact, for the Hamiltonian (\ref{4-4}%
), the S-matrix coefficients read (see \cite{Laura 1997}) 
\begin{equation}
S(\omega )=1-2\pi i\langle \omega |V|\omega \rangle -2\pi i\langle \omega |V%
\frac{1}{\omega +i0-H}V|\omega \rangle  \label{4-7}
\end{equation}
Then, if $V$ is well behaved, i.e. the analytical continuation of $V|\omega
\rangle $ is a vector value analytical function \cite{Kato}, functions $%
R(\omega )$ and $S(\omega )$ have the same poles.\footnote{%
If $V$ is not well behaved, some new poles may appear.}

As before, for the sake of simplicity we will assume that $R(z)$ has just
one pole $z_{0}$,\footnote{$z_{0}$ is the pole of the $|\psi (t)\rangle
-evolution$, $Z_{0}=\overline{z}_{0}-z_{0}$ is the pole of the $\rho
(t)-evolution$. The minus sign in the $-z_{0}$ produces the change from the
lower half-plane for the decaying processes in the $|\psi (t)\rangle
-evolution$ to the upper half-plane for these processes in the $\rho
(t)-evolution$.} and we will only consider pure states. On this basis we
will show that, if the continuation to the lower second sheet of $\langle
\varphi |\omega \rangle $ has no poles, the continuation to the lower second
sheet of $\langle \varphi |\omega \rangle ^{+}$ gets the pole $z_{0}$ of
function $R(z)$. In fact, from the Lippmann-Schwinger equation we know that
there are two eigenbasis for $H$ (see \cite{Laura 1997}), 
\begin{equation}
|\omega \rangle ^{\pm }=|\omega \rangle +\frac{1}{\omega \pm i0-H}V|\omega
\rangle  \label{4-8}
\end{equation}
where $\pm i0$ symbolizes the analytical continuation to the lower (upper)
half-plane in the second sheet. As explained, we will consider only the
decaying case and, therefore, we will only use $\{|\omega \rangle ^{+}\}$.
Then, 
\begin{equation}
\langle \varphi |\omega \rangle ^{+}=\langle \varphi |\omega \rangle
+\langle \varphi |\frac{1}{\omega +i0-H}V|\omega \rangle  \label{4-9}
\end{equation}
and we can make the analytical continuation of this equation to the lower
second sheet 
\begin{equation}
\langle \varphi |z\rangle ^{+}=\langle \varphi |z\rangle +\langle \varphi |%
\frac{1}{z-H}V|z\rangle  \label{4-10}
\end{equation}
If we suppose, as before, that $V$ is a well behaved vector value function,
then even if $\langle \varphi |\omega \rangle $ has no poles (as required in
Case 1), $\langle \varphi |\omega \rangle ^{+}$ has a pole at $z_{0}$,
coming from the resolvent term. The same applies to $^{+}\langle z|\psi
\rangle $: even if $\langle z|\psi \rangle $ has no poles, $^{+}\langle
z|\psi \rangle $ gets the poles of the resolvent. The argument can be
extended to states and observables: even if $\langle z|\rho |z^{\prime
}\rangle $ has no poles, $^{+}\langle z|\rho |z^{\prime }\rangle ^{+}$ has
poles, and even if $\langle z|O|z^{\prime }\rangle $ has no poles, $%
^{+}\langle z|O|z^{\prime }\rangle ^{+}$ has poles.

As a consequence, if the initial condition of the 'interaction' period is
given by the states and operators of a previous 'free' period which, as
shown in Case 1, have no poles in the eigenbasis of $H_{0}$ (precisely, the
continuations of $\rho (E,E^{\prime })=\langle E|\rho |E^{\prime }\rangle $
and $O(E,E^{\prime })$ $=\langle E|O|E^{\prime }\rangle $ have no poles),
then $\rho (z,z^{\prime })=\langle z|\rho |z^{\prime }\rangle $ and $%
O(z,z^{\prime })=\langle z|O|z^{\prime }\rangle $ have no poles but $%
^{+}\langle z|\rho |z^{\prime }\rangle ^{+}$ and $^{+}\langle z|O|z^{\prime
}\rangle ^{+}$ {\it do have poles} that produce the dumping factor of eq.(%
\ref{3-5}).\footnote{%
Moreover, if conditions 1 or 2 of Case 1 are not satisfied, $\rho
(z,z^{\prime })=\langle z|\rho |z^{\prime }\rangle $ and $O(z,z^{\prime
})=\langle z|O|z^{\prime }\rangle $ may have poles, as it will be shown in
Section VII. Nevertheless, in certain sense more poles are welcomed because
what we are essentially trying to prove is that decoherence time is very
small.} These results are presented in great detail in paper \cite{Liotta},
where the dumping factor appears in eq.(70), precisely, 
\[
(\rho (t)|O)=\int_{0}^{\infty }d\omega (\rho _{0}|\Phi _{\omega })(%
\widetilde{\Phi }_{\omega }|O)+e^{i(\overline{z}_{0}-z_{0})t}(\rho _{0}|\Phi
_{00})(\widetilde{\Phi }_{00}|O) 
\]
\[
+\int_{\Gamma }dz^{\prime }e^{i(\overline{z}_{0}-z^{\prime })t}(\rho
_{0}|\Phi _{0z^{\prime }})(\widetilde{\Phi }_{0z^{\prime }}|O)+\int_{%
\overline{\Gamma }}dze^{i(z-z_{0})t}(\rho _{0}|\Phi _{z0})(\widetilde{\Phi }%
_{z0}|O) 
\]
\begin{equation}
+\int_{\Gamma }dz^{\prime }\int_{\overline{\Gamma }}dze^{i(z-z^{\prime
})t}(\rho _{0}|\Phi _{zz^{\prime }})(\widetilde{\Phi }_{zz^{\prime }}|O)
\label{4-11}
\end{equation}
Independently of the precise meaning of each symbol (which can be found in 
\cite{Liotta}), it is quite clear that the term $e^{i(\overline{z}%
_{0}-z_{0})t}(\rho _{0}|\Phi _{00})(\widetilde{\Phi }_{00}|O)=e^{iZ_{0}t}(%
\rho _{0}|\Phi _{00})(\widetilde{\Phi }_{00}|O)$ represents the main
contribution to decoherence, that is, the pole-contribution. The first term
is the 'constant term' and the last three terms are background terms.%
\footnote{%
All this computation can also be made by using the Laplace transform (see 
\cite{Grecos}) 
\[
\exp (-iLt)=\frac{1}{2\pi i}\int_{C}\exp (-izt)\frac{1}{L-z}dz 
\]
with the same result. In this case, all the conditions required in this
section are also needed.} Furthermore, it can be proved that a sufficient
condition for the coefficient $(\rho _{0}|\Phi _{00})(\widetilde{\Phi }%
_{00}|O)$ be well defined is that $\rho (z,z^{\prime })=\langle z|\rho
|z^{\prime }\rangle $ and $O(z,z^{\prime })=\langle z|O|z^{\prime }\rangle $
have no poles.

\section{Decoherence time for microscopic systems}

In these three final sections we will present some estimates of decoherence
time, in order to show that self-induced decoherence can account for already
known results and opens the way to more detailed models.

As already explained, if $\gamma $ is the imaginary part of the pole (or of
the pole closer to the real axis), the decoherence time is 
\begin{equation}
t_{D}=\frac{\hbar }{\gamma }  \label{5-1}
\end{equation}
because, as we have said, the {\it characteristic decaying time of the
fluctuating term} of $\langle O\rangle _{\rho (t)}=(\rho (t)|O)$ is the {\it %
decoherence time} $t_{D}$.

The decoherence time can be estimated in particular cases like e.g. the
Friedrich model studied in papers \cite{Cast-Laura 1997} and \cite{Rolo},
where we obtain 
\begin{equation}
t_{D}=\frac{\hbar }{2\pi |V_{\Omega }|^{2}}  \label{5-2}
\end{equation}
being $V_{\Omega }$ the interaction function. It is clear that, if the
interaction vanishes, $t_{D}\rightarrow \infty $. In turn, if the
characteristic energy $2\pi |V_{\Omega }|^{2}\sim V$ is, say, $1$
electron-volt (a natural energy scale for quantum atomic interactions, see
e.g. \cite{Kuyatt}), the decoherence time is $\sim 10^{-15}s$.\footnote{%
This is, of course, a general case: $\gamma $ is usually of the order of
magnitude of the characteristic interaction energy $V$.}

This means that, when the theory is calibrated in such a way that the 'free'
period does not lead to decoherence, a generic system does decohere (and, in
general, {\it very fast}) in the 'interaction' period.

It is interesting to remark that the method of paper \cite{Liotta} was
compared with the usual methods of nuclear physics (see \cite{Zeldovich}, 
\cite{Gyarmati}, \cite{Simon}) in the case of a $^{208}P_{b}(2d_{5/2})$
proton state in a Woods-Saxon potential, including spin-orbits interaction
with parameters as in paper \cite{Curuchet}, with an excellent agreement
(see fig.3 of \cite{Liotta}). In this case, $\gamma =10^{-1}Mev$ and $%
t_{D}\sim 10^{-20}s$.

\section{Decoherence time of macroscopic bodies}

A pure state of a macroscopic object $|\psi \rangle $ can be considered as
the tensor product of $N$ states $|\psi _{i}\rangle $ of microscopic
particles 
\begin{equation}
|\psi \rangle =\bigotimes_{i=1}^{N}|\psi _{i}\rangle  \label{6-2}
\end{equation}
A generic total Hamiltonian reads 
\begin{equation}
H=\sum_{i=1}^{N}H_{i}+\sum_{i=1}^{N}V_{i}+\sum_{i,j=1}^{N}V_{ij}+%
\sum_{i,j,k=1}^{N}V_{ijk}+...  \label{6-3}
\end{equation}
where the $H_{i}$ are the free Hamiltonians of the microscopic particles,
the $V_{i}$ are the averages of the interactions between each particle and
the remaining particles, the $V_{ij}$ are the two particle interactions, and
so forth. The eigenvectors of $H$ are $|\omega ,x_{1},...,x_{N-1}\rangle $,
such that 
\begin{equation}
H|\omega ,x_{1},...,x_{N-1}\rangle =\omega |\omega ,x_{1},...,x_{N-1}\rangle
\label{6-4}
\end{equation}
where $\omega $ is the {\it total energy} and $x_{1},...,x_{N-1}$ the
remaining labels necessary to define the eigenstate. Since decoherence
occurs in variable $\omega $, we can ignore the rest of the labels for our
argument. Disregarding for simplicity the $V_{ij},V_{ijk},...$, and
considering all the $V_{i}$ equal, we obtain 
\begin{equation}
V=\sum_{i=1}^{N}V_{i}=NV_{i}  \label{6-5}
\end{equation}
Then, the characteristic energy in eq.(\ref{5-2}) is now $NV_{i}$. Let us
suppose again that all the $V_{i}$ are of the order of $1$ electron-volt. If
we consider a macroscopic object of one mol, where $N=10^{24}$, the
decoherence time results 
\[
t_{D}=10^{-39}s 
\]
a very tiny time indeed. This decoherence time is so close to Plank time $%
10^{-43}s$ that it should be considered more as an illustration than as a
physical result. Nevertheless, it shows that decoherence is fantastically
fast in macroscopic bodies.

\section{Initial thermal bath}

Finally, we will consider the case of a system with a thermal bath,
following the formalism of paper \cite{Laura-Cast 1998-E}, section IV.B. For
the model considered in that section, an oscillator in a thermal bath, the
fluctuating term of eq.(\ref{2-7}) is 
\begin{equation}
\int O_{{\bf pp}^{\prime }}\overline{\rho }_{{\bf pp}^{\prime }}\,e^{i\frac{%
(\omega _{p}-\omega _{p^{\prime }})t}{\hbar }}d{\bf p}d{\bf p}^{\prime }
\label{7-1}
\end{equation}
where $\overline{\rho }_{{\bf pp}^{\prime }}=(\rho |A_{{\bf p}}^{\dagger }A_{%
{\bf p}^{\prime }})$, given in eq.(37) of \cite{Laura-Cast 1998-E}, is the
initial condition of the oscillator and thermal bath, 
\[
(\rho |A_{{\bf p}}^{\dagger }A_{{\bf p}^{\prime }})=\rho ({\bf p})\delta
^{3}({\bf p-p}^{\prime })+\frac{\langle n\rangle _{0}V_{p}V_{p^{\prime }}}{%
\eta _{+}(\omega _{p})\eta _{-}(\omega _{p^{\prime }})}+\frac{%
V_{p}V_{p^{\prime }}\rho ({\bf p})}{\eta _{-}(\omega _{p^{\prime }})(\omega
_{p^{\prime }}-\omega _{p}-i0)} 
\]
\begin{equation}
+\frac{V_{p}V_{p^{\prime }}\rho ({\bf p})}{\eta _{+}(\omega _{p})(\omega
_{p}-\omega _{p^{\prime }}+i0)}+\frac{V_{p}V_{p^{\prime }}}{\eta _{+}(\omega
_{p})\eta _{-}(\omega _{p^{\prime }})}\int \frac{d{\bf k}V_{k}^{2}\rho ({\bf %
k})}{\eta _{-}(\omega _{p^{\prime }})(\omega _{p^{\prime }}-\omega
_{p}-i0)(\omega _{p}-\omega _{p^{\prime }}+i0)}  \label{7-2}
\end{equation}
where $\rho ({\bf p})$ is given in eq.(40) of that paper, 
\begin{equation}
\rho ({\bf p})=\frac{1}{e^{\beta \omega _{p}}-1}  \label{7-3}
\end{equation}
and where $\beta =1/kT$. Again, independently of the precise meaning of each
term (which can be found in \cite{Laura-Cast 1998-E}), it is obvious that
the $\rho ({\bf p})$ factor has a pole in the $\omega _{p}$ complex plane
and, therefore, the initial condition is not pole-free. Then, defining as
before 
\begin{equation}
\nu =\omega _{p}-\omega _{p^{\prime }},\qquad 2\lambda =\omega _{p}+\omega
_{p^{\prime }},\qquad \omega _{p}=\lambda +\frac{\nu }{2},\qquad etc.
\label{7-4}
\end{equation}
the terms of eq.(\ref{7-1}) have the form 
\begin{equation}
\int ...\frac{1}{e^{\beta (\lambda +\frac{\nu }{2})}-1}e^{i\frac{\nu t}{%
\hbar }}d\lambda d\nu  \label{7-5}
\end{equation}
where the dots symbolize other factors coming from the interaction (which
may have poles not considered here). When we introduce the complex variable $%
z=\omega +i\gamma $, the poles of the factor $\rho ({\bf p})$ turn out to be
located where the following equation is satisfied 
\begin{equation}
e^{\beta (\lambda +\frac{z}{2})}=e^{\beta (\lambda +\frac{\omega }{2}%
)}\left( \cos \frac{\gamma }{2kt}+i\sin \frac{\gamma }{2kt}\right) =1
\label{7-6}
\end{equation}
Then, the poles are located in the coordinates 
\begin{equation}
\omega =-2\lambda ,\qquad \gamma =4\pi nkT  \label{7-7}
\end{equation}
where $n$ is an integer number. For $n=1$ we obtain the decoherence time 
\begin{equation}
t_{D}=\frac{\hbar }{4\pi kT}  \label{7-8}
\end{equation}
which, for room temperature, $T\sim 10^{2}{{}^{\circ }}K$, gives $%
t_{D}=10^{-13}s$ for a single particle system, and $t_{D}=10^{-37}s$ for a
mol-particle system.

Moreover, if $S_{1}=\hbar $ is the characteristic action for a particle
system,\footnote{%
E.g.:
\par
1.- In the harmonic oscillator, the unidimensional coordinates are $%
q=(M\omega /\hbar )^{1/2}Q$ and $p=(1/M\omega \hbar )^{1/2}P$; then, by
making $q=p=1$ we obtain $S_{1}=QP=\hbar $,
\par
2.- In a model with spherical symmetry we have $L_{z}Y(\theta ,\varphi
)=m\hbar Y(\theta ,\varphi )$; then, by making $m=1$ we obtain $%
S_{1}=\varphi L_{z}=\hbar $,
\par
and so forth.} and $S=ML^{2}/\Upsilon $ is the action of a macroscopic
system (where $M$, $L$ and $\Upsilon $ are the characteristic mass, length
and time), the particle number can be estimated as 
\begin{equation}
N=\frac{S}{S_{1}}=\frac{ML^{2}}{\hbar \Upsilon }  \label{7-9}
\end{equation}
and the decoherence time reads 
\begin{equation}
t_{D}\sim \Upsilon \frac{\hbar ^{2}}{ML^{2}kT}  \label{7-10}
\end{equation}

In the particular case that $\Upsilon =\gamma _{0}^{-1}$, $L=L_{0}$ and $M=M$
are the characteristic time, length and mass of the model of Section 4.1 of
paper \cite{Paz}, when we introduce the de Broglie length $\lambda _{DB}=%
\frac{\hbar }{\sqrt{3MkT}}\sim \frac{\hbar }{\sqrt{MkT}}$, we obtain 
\begin{equation}
t_{D}\sim \gamma _{0}^{-1}\left( \frac{\lambda _{DB}}{L_{0}}\right) ^{2}
\label{7-11}
\end{equation}
namely, eq.(4.10) of paper \cite{Paz} which represents the decoherence time
for the model of section 4.1. This shows the agreement between the
self-induced and the einselection approaches in the obtained value of the
decoherence time.

\section{Conclusion}

In a series of previous papers we have developed an approach to decoherence
that avoids the drawbacks of the einselection approach. In the present paper
we have shown that our formalism supplies a precise method for computing the
decoherence time, and that the results obtained with such a method are
physically meaningful and coincide with those obtained in the literature.

However, we are aware of the fact that a great number of results have been
obtained in the context of the einselection program when compared with the
cases treated by means of the self-induced approach. Therefore, our future
work has to be directed to enlarge the set of applications of our theory. We
consider that this task will be worth the effort to the extent that
decoherence is a key element in the explanation of the emergence of
classicality from the quantum world.

\appendix

\section{Hartog theorem}

A necessary mathematical remark is in order: from the Hartog theorem, we
know that the realm of analytical functions of more than one complex
variable is much more involved than that of just one variable. In papers 
\cite{Cast-Laura 1997}, \cite{Rolo} and \cite{Liotta}, we have considered
the analytical continuation of two variables, $z$ and $z^{\prime }$, in
products like $f_{1}(z)$ $f_{2}(z^{\prime })$; this suggests that we should
have worked with the theory of analytical functions of two complex
variables. However, this is not the case because the only relevant variable
for the problems considered in the quoted papers is the difference $%
Z=z-z^{\prime }$ that appears in the evolution factor $e^{i(z-z^{\prime })t}$%
; then, we can always introduce a change of variables 
\[
z+z^{\prime }=2\lambda ,\qquad z-z^{\prime }=Z 
\]
\begin{equation}
z=\lambda +\frac{Z}{2},\qquad z^{\prime }=\lambda -\frac{Z}{2}  \label{A-1}
\end{equation}
and functions $f(z,z^{\prime })$, like the product $f_{1}(z)$ $%
f_{2}(z^{\prime })$, have to be considered as functions $f(z,z^{\prime
})=f(\lambda +\frac{Z}{2},\lambda -\frac{Z}{2})$ where $Z\in {\Bbb C}$; but
since in all the cases we have taken $\lambda \in {\Bbb R}$, $f(z,z^{\prime
})$ is really a function of only one complex variable $Z$.

For instance, in eq. (71) of paper \cite{Liotta}, the symbol $cont_{\omega
\rightarrow \overline{z}_{0}}cont_{\omega ^{\prime }\rightarrow z_{0}}$
should be understood as 
\begin{equation}
cont_{\lambda +\frac{\nu }{2}\rightarrow \widetilde{\omega }+i\frac{\gamma }{%
2}}cont_{\lambda -\frac{\nu }{2}\rightarrow \widetilde{\omega }-i\frac{%
\gamma }{2}}=cont_{\lambda \rightarrow \widetilde{\omega }}cont_{\nu
\rightarrow i\gamma }  \label{A-2}
\end{equation}
where $\lambda $ is a real number. In this equation, only the second
continuation of the r.h.s. is an analytical continuation in the complex
plane, being the first one a simple change of a real variable, from $\lambda 
$ to $\widetilde{\omega }$.

These considerations show that papers \cite{Cast-Laura 1997}, \cite{Rolo}, 
\cite{Liotta} and, in general, the method presented in this paper, are free
from Hartog's objection.

\section{Two-times evolution}

Let us now generalize eq.(\ref{4-4}) by adding two interactions 
\begin{equation}
H=H_{0}+V=\int_{0}^{\infty }d\omega \omega |\omega \rangle \langle \omega
|+\int_{0}^{\infty }\int_{0}^{\infty }[V^{(1)}(\omega ,\omega ^{\prime
})+V^{(2)}(\omega ,\omega ^{\prime })]|\omega \rangle \langle \omega
^{\prime }|d\omega d\omega ^{\prime }  \label{B-1}
\end{equation}
where $V^{(1)}$represents a macroscopic interaction and $V^{(2)}$ represents
a microscopic interaction: $V^{(1)}(\omega ,\omega ^{\prime })\gg
V^{(2)}(\omega ,\omega ^{\prime })$. As a consequence, in a first step we
can neglect $V^{(2)}(\omega ,\omega ^{\prime })$ and repeat what was said in
Section IV. Then, we can begin with considering a Hamiltonian 
\begin{equation}
H^{(1)}=H_{0}+V_{1}=\int_{0}^{\infty }\omega |\omega \rangle \langle \omega
|d\omega +\int_{0}^{\infty }\int_{0}^{\infty }V^{(1)}(\omega ,\omega
^{\prime })|\omega \rangle \langle \omega ^{\prime }|d\omega d\omega
^{\prime }  \label{B-2}
\end{equation}
If we change the basis to $\{|\omega \rangle _{(1)}^{+}\}$, we obtain 
\begin{equation}
|\omega \rangle _{(1)}^{+}=|\omega \rangle +\frac{1}{\omega +i0-H}%
V^{(1)}|\omega \rangle  \label{B-3}
\end{equation}
Since the interaction $V^{(1)}$ is macroscopic, the dumping of the
off-diagonal terms has a characteristic time of $10^{-37}s-10^{-39}s$.
Therefore, after an initial period much larger than this magnitude, the
state can be considered nearly diagonal for all practical purposes.

However, the state is not yet in complete equilibrium, because the
interaction $V^{(2)}$ is always present: now it becomes relevant. Then,
after the initial period we can consider the total Hamiltonian as 
\[
H=H^{(1)}+V_{2}=\int_{0}^{\infty }\omega |\omega \rangle _{(1)}^{+}\langle
\omega |_{(1)}^{+}d\omega +\int_{0}^{\infty }\int_{0}^{\infty }V^{(2)\prime
}(\omega ,\omega ^{\prime })||\omega \rangle _{(1)}^{+}\langle \omega
^{\prime }|_{(1)}^{+}d\omega d\omega ^{\prime } 
\]
where $V^{(2)\prime }(\omega ,\omega ^{\prime })$ is $V^{(2)}(\omega ,\omega
^{\prime })$ in the new basis $\{|\omega \rangle _{(1)}^{+}\}$. When we make
a final change of basis 
\[
|\omega \rangle ^{+}=|\omega \rangle _{(1)}^{+}+\frac{1}{\omega +i0-H}%
V^{(2)}|\omega \rangle _{(1)}^{+} 
\]
we can compute the characteristic time in this case. But now this time may
be of the order of $1s$, that is, the relaxation time of a macroscopic body.
In this way we can describe a two-times process, with a extremely short
decoherence time and a long relaxation time.

\end{document}